\documentclass[prl,twocolumn,superscriptaddress,showpacs,floatfix]{revtex4-1}
\usepackage{amsfonts}
\usepackage{amsmath}
\usepackage{amssymb}
\usepackage{graphicx}

\begin{document}

\title{Generation of maximally correlated states in the absence of entanglement}
\date{\today}

\author{C. E. L\'opez}
\affiliation{Departamento de F\'{\i}sica, Universidad de Santiago de Chile, USACH, Casilla 307 Correo 2 Santiago, Chile}
\author{F. Albarr\'an-Arriagada}
\affiliation{Departamento de F\'{\i}sica, Universidad de Santiago de Chile, USACH, Casilla 307 Correo 2 Santiago, Chile}
\author{S. Allende}
\affiliation{Departamento de F\'{\i}sica, Universidad de Santiago de Chile, USACH, Casilla 307 Correo 2 Santiago, Chile}
\affiliation{Center for the Development of Nanoscience and Nanotechnology, 9170124, Estaci\'on Central, Santiago, Chile}
\author{J. C. Retamal}
\affiliation{Departamento de F\'{\i}sica, Universidad de Santiago de Chile, USACH, Casilla 307 Correo 2 Santiago, Chile}
\affiliation{Center for the Development of Nanoscience and Nanotechnology, 9170124, Estaci\'on Central, Santiago, Chile}

\begin{abstract}
We study the generation of maximally correlated states of two qubits in the absence of quantum entanglement. We show that stationary maximally correlated states can be generated under the assistance of a  collective dissipative dynamics. The absence of entanglement necessarily requires maximal entanglement to an environment. The conditions under which two qubits can be maximally correlated to a finite environment are studied. We find the existence of maximally correlated states without entanglement  for $3\otimes 3$ bipartite quantum states.
\end{abstract}

\maketitle

\section{introduction}
Entanglement in multipartite quantum systems is a central issue in Quantum Information and Quantum Computation, and  has received considerable attention from a  theoretical \cite{horo1,Geza} and an experimental side \cite{brasil1,brasil2,laurat}. 
Today we know that there exist other correlations than entanglement which can be embodied in the quantum state of two qubits~\cite{review}. From an entropic point of view, the total amount of  correlations, quantum and classical,  can be assumed to be described by the total quantum mutual information. The total quantum correlations are obtained after subtraction of   classical correlations from quantum mutual information. The operational definition of classical correlations at the quantum level, lead us with the definition of { \it quantum discord} ($Q$)~\cite{discord}, a measure of quantum correlations. Quantum correlations could be geometrically understood as a distance, in relative entropy, from a given quantum state to the closest classical state,  differentiating explicitly the entanglement defined as the distance, in relative entropy, from a given quantum state to the closest separable state~\cite{modi}. In the absence of entanglement a residual correlation can still be present in a separable quantum state given by the distance, in relative entropy, from the separable state to the closest classical state. This is the so called { \it quantum dissonance}. Thus quantum dissonance is the correlation existing in the absence of entanglement. The research in this field has received a considerable attention since the discovery that certain computational task could be accomplished in the absence of entanglement \cite{DQC1,lroa}. A prolific  of scientific work has been dedicated to  characterize, measure and the generation of states having correlations other than entanglement. 

Elucidate whether or not there is a physical process under which correlated quantum states of two qubits could be created in the absence of entanglement is a fundamental problem we address in this work.  In particular, the existence of a protocol to create states having a maximal amount of { \it quantum dissonance}. The availability of such states should be important to accomplish computational task,  not requiring entanglement, in quantum systems for which the availability of quantum coherent control could be a limited physical resource. Finally we find numerical evidence  that  maximally correlated quantum states could be created for bipartite higher dimensional quantum systems in the absence of entanglement.

\section{Maximally correlated qubits with no entanglement}

The class of states we are attempting to generate have to be found within the set of separable states. In particular it has been defined the class of maximally discordant mixed states, which  is a class of rank 2 and 3 states having maximal quantum discord versus classical correlations. The class of states thus defined are given by \cite{galve}:
\begin{equation}
\rho=\epsilon\mid\Phi^{+}\rangle \langle \Phi^{+}|+(1-\epsilon)(m |01 \rangle \langle 01| +(1-m)| 10 \rangle \langle 10|)
\label{rank3}
\end{equation}
where $|\Phi^{+}\rangle = (|00\rangle+|11\rangle)/\sqrt(2)$. For these states the entanglement is readily found to be $E=\max\{0, \epsilon-2(1-\epsilon)\sqrt{m(1-m)}\}$. The class of separable states are those for which  $\epsilon$ is below the border of separability given by $\epsilon_s=2\sqrt{m(1-m)}/(1+2\sqrt{m(1-m)})$. Given that $m\in [0,1]$, $\epsilon_s(m)$ is symmetric around $m=1/2$ reaching the maximum value $\epsilon_s |_{max}=1/2$. All these states belong to the class of dissonant states, that is the quantum correlations embodied in such states is only dissonance \cite{galve}. Let us briefly remind the main concepts involved.  Quantum correlations are defined as the difference between {\it quantum mutual information} $\cal{I}=S(\rho_A)+S(_B)-S(\rho_{AB})$ and the {\it classical correlations} $C(\rho_{AB})=\max{\{S(\rho_{A})-S(\rho_{A|B})\}}$, where $S(\rho_{A|B})$ is the conditional entropy of $A$ given a measurement on system $B$ and the optimization is over all possible projective measurement on system $B$. The conditional entropy for  the case we are interested  can be  calculated using the Ali-Rau-Alber results \cite{ali}.

It is not difficult to find that for $m=1/2$ and $\epsilon=1/3$ the quantum mutual information is maximized and the classical correlations, minimized. That is, the incoherent superposition of three Bell states
\begin{equation}
\rho=\frac{1}{3}(\mid\Phi^{+}\rangle \langle \Phi^{+}|+ |01 \rangle \langle 01| +| 10 \rangle \langle 10|),
\label{max1}
\end{equation}
correspond to a maximally dissonant state. Other realization could be obtained by  changing subpaces:
\begin{equation}
\rho_{\rm }=\frac{1}{3}(\mid\Psi^{+}\rangle \langle \Psi^{+}|+ |00 \rangle \langle 00| +| 11 \rangle \langle 11|)
\label{maximal}
\end{equation}
where $|\Psi^{+}\rangle = (| 01\rangle+|10\rangle)/\sqrt{2}$. 
\subsection{Reservoir assisted generation}
The important issue we address in this work is the search for a dynamical process that could allow us to generate this class of quantum states without generating entanglement. From the state above, we guess that such dynamics should be conditioned within a portion of  the Hilbert space such that only three Bell states are involved. We will use the notation:
\begin{eqnarray*}
|1\rangle&=& \frac{1}{\sqrt{2}}(|00\rangle+|11\rangle) , \quad |2\rangle= \frac{1}{\sqrt{2}}(|00\rangle-|11\rangle)\nonumber \\
|3\rangle&=& \frac{1}{\sqrt{2}}(|01\rangle+|10\rangle), \quad |4\rangle= \frac{1}{\sqrt{2}}(|01\rangle-|10\rangle)\nonumber 
\end{eqnarray*}
Suppose we have a physical process that only involves the subspace $\{|1\rangle,|2\rangle,|3\rangle\}$ such that the probability amplitude evolves according to:
\begin{eqnarray*}
\dot{\rho}_{11}&=&-3\gamma \rho_{11}+\gamma \nonumber \\
\dot{\rho}_{22}&=&-3\gamma \rho_{22}+\gamma \nonumber \\
\dot{\rho}_{33}&=&-3\gamma \rho_{33}+\gamma \nonumber 
\end{eqnarray*}
where $\gamma$ is a constant. It is not difficult to see that  such equations evolves towards an stationary maximally dissonant state. The existence of this dynamics is conditioned to the restriction that no coherence is created among different Bell states, and that the initial probability amplitude in state $|4\rangle$ is zero. Under such restriction the system of equations above could be written as:
\begin{eqnarray}
\dot{\rho}_{11}&=&-2\gamma \rho_{11}+\gamma (\rho_{22}+\rho_{33}) \nonumber \\
\dot{\rho}_{22}&=&-2\gamma \rho_{22}+\gamma (\rho_{11}+\rho_{33})\\
\dot{\rho}_{33}&=&-2\gamma \rho_{33}+\gamma (\rho_{11}+\rho_{22})\nonumber 
\label{pro123}
\end{eqnarray}
Can this system of equations be originated from a dynamics of the Lindblad form? If this is the case, the selected Bell states should be eigenstates of the operators included in the master equation, and the transitions between Bell induced by the Lindblad form should be restricted to this subspace. For completeness, it is no difficult to verify that any local Lindblad ${\cal{L}}_\gamma(\rho)=-\gamma[1\otimes A ,[1\otimes A,\rho]]$ with $A=\sigma, \sigma^+, \sigma_x,\sigma_y, \sigma_z$ cannot lead to Eqs. (\ref{pro123}). 

\begin{figure}
\includegraphics[width=50mm]{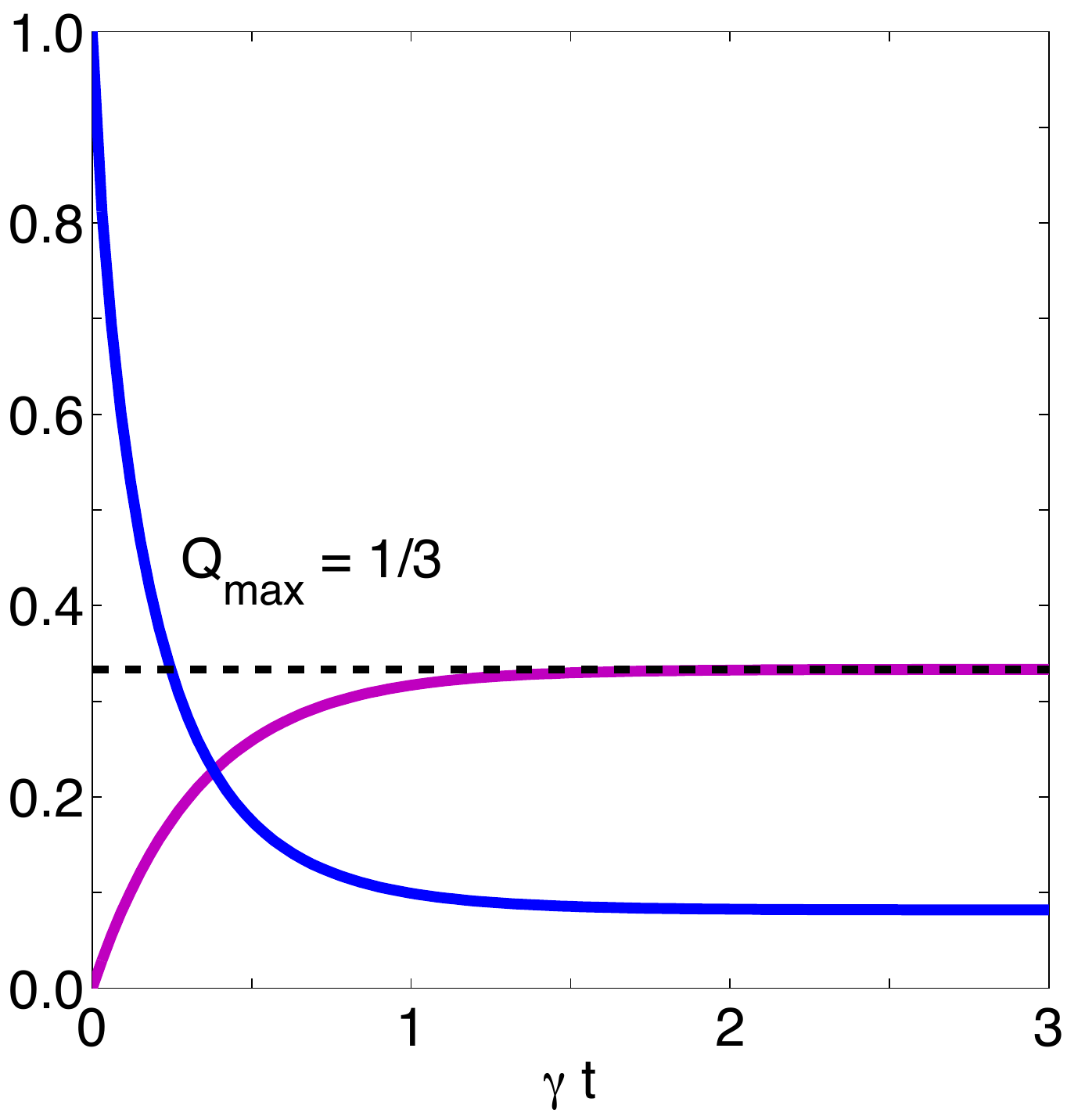} \caption{Evolution of discord $Q(t)$ (purple line) and classical correlations $C(t)$ (blue line).}
\label{fig1}  
\end{figure}

Since the Bell states $\{|1\rangle,|2\rangle,|3\rangle \}$ are spanned by the symmetric spin triplet, the Lindblad form must necessarily include collective spin operators. One first attempt could be to consider a collective spin decay described by the operator $S=\sigma_1+\sigma_2$. This operator connect states $|1\rangle$ and $|3\rangle$, but state $|3\rangle$ is connected through $S^{\dagger}$ back to a combination of $|1\rangle$ and $|2\rangle$, making this operator not suitable to lead to Eqs. (\ref{pro123}). Alternatively we can consider the cartesian component of collective spin operators:
\begin{eqnarray}
S_x&=&\frac{\sigma_{1x}+\sigma_{2x}}{2}, \nonumber \\ S_y&=&\frac{\sigma_{1y}+\sigma_{2y}}{2}, \\S_z&=&\frac{\sigma_{1z}+\sigma_{2z}}{2}\nonumber
\end{eqnarray}
As can be directly checked, these operators have matrix elements between Bell states in the symmetric subspace given by:
\begin{eqnarray}
S_x|1\rangle&=&|3\rangle ,\,S_x|3\rangle =|1\rangle, \,\,S_x|2\rangle =0\nonumber \\ 
S_y|1\rangle&=&0,\, S_y|2\rangle=-i|3\rangle,\, S_y|3\rangle=i|2\rangle \\
S_z|1\rangle&=&|2\rangle,\, S_z|2\rangle=|1\rangle,\, S_z|3\rangle\nonumber =0.
\end{eqnarray}
They have the right matrix elements that could give rise to (\ref{pro123}) when the system evolves under Lindblad form
\begin{equation}
{\cal{L}}\rho = {\cal{L}}_\gamma (S_x)\rho+{\cal{L}}_\gamma (S_y)\rho+{\cal{L}}_\gamma (S_z)\rho
\label{lindblad}
\end{equation}
A more general situation would be that of a system evolving under a free hamiltonian $H_0$ and a Lindblad  with different decay rates such as:
\begin{equation}
\dot{\rho}=-i[H_0,\rho]+{\cal{L}}_{\gamma_x} (S_x)\rho+{\cal{L}}_{\gamma_y} (S_y)\rho+{\cal{L}}_{\gamma_z} (S_z)\rho
\label{total}
\end{equation}
Assuming that Bell states in the symmetric subspace are eigenstates of $H_0$, the general equations for the matrix elements of $\rho$ are given by,
\begin{eqnarray}
\dot{\rho}_{11}&=&\gamma_x (\rho_{33}-\rho_{11})+\gamma_z (\rho_{22}-\rho_{11}), \nonumber \\
\dot{\rho}_{22}&=&\gamma_y (\rho_{33}-\rho_{22})+\gamma_z (\rho_{11}-\rho_{22}),\nonumber \\
\dot{\rho}_{33}&=&\gamma_x (\rho_{11}-\rho_{33})+\gamma_y (\rho_{22}-\rho_{33}), 
\label{pro123}
\end{eqnarray}
provided that state $|4\rangle$ is not initially populated. This is the case when considering and initially unentangled state $|00\rangle$ which in the Bell basis can be written as $(|1\rangle+|2\rangle)/\sqrt{2}$. In this situation we use the normalization condition $\rho_{33}=1-\rho_{22}-\rho_{11}$ and then the rate equations become
\begin{eqnarray}
\dot{\rho}_{11}&=&-(2\gamma_x +\gamma_z)\rho_{11}+(\gamma_z -\gamma_x)\rho_{22}+\gamma_x \nonumber \\
\dot{\rho}_{22}&=&-(2\gamma_y +\gamma_z)\rho_{22}+(\gamma_z -\gamma_y)\rho_{11}+\gamma_y 
\label{pro1234}
\end{eqnarray}
For example, when $\gamma_x =\gamma_y =\gamma$ and $\gamma_z = 0$, the solution has the following form
\begin{equation}
\rho_{11}(t)=\rho_{22} (t)= \frac{1}{3} \bigg	\{ 1+\frac{1}{2}e^{-3 \gamma t}\bigg \},
\end{equation}
that is, after a long time: $\rho_{11} = \rho_{22} = 1/3$ which corresponds to the steady maximally dissonant state~(\ref{maximal}). This is also true whenever one of the decay rates $(\gamma_x , \gamma_y , \gamma_z )$ is zero and the other two are different from zero. In Fig.~\ref{fig1}, where the evolution of quantum discord $Q(t)$ and classical correlations $C(t)$ are shown as a function of $\gamma t$ we observe that discord reaches and stationary maximum value of $Q =1/3$ when $\gamma t\rightarrow \infty$. 

It is interesting to know which are quantum resources  needed to generate this class of quantum states. In previous analysis  we concluded that maximally dissonant states are generated throughout an open dynamics involving  collective spin operators. Whether or not such open dynamics could be generated in the absence of quantum entanglement is the key question. A physical realization for such dynamics is not usual to be found in quantum optical systems. However, as developed in reference \cite{sipe}, such dynamics appears for a pair of impurity-bound electrons interacting with a bath of conduction band electrons in a semiconductor. In such case the electrons in the conduction band produce and RKKY interaction between localized spins and an effective dissipative structure as in Eq.~(\ref{lindblad}). It has recently been found that a dynamics with $\gamma_y=0$ can be obtained for two qubits subject to independent noisy classical fields, that  leads to maximally dissonant states as shown in ref.~\cite{altintas1}.

The physical architecture behind the open dynamics in equation (\ref{total}) actually needs the presence of an intermediary quantum system, for instance the electrons in the conduction band, whose degrees of freedom are traced out. This means that even when no actual entanglement is being generated between localized spins, entanglement has to be present in the interaction between localized qubits with the electrons in the conduction band. In order to clarify this, we consider the state in Eq.~(\ref{rank3}) and write its purified version for $m=1/2$ and $\epsilon = 1/3$,
\begin{equation}
|\Psi \rangle =\frac{1}{\sqrt{3}} \left( | \Psi^{+}\rangle |e_1 \rangle + |00\rangle |e_2 \rangle  + | 11 \rangle |e_3 \rangle \right)
\end{equation}
as we see, the maximally dissonant two-spin state is maximally entangled with the purification state space.  Notice that both, the two spins and the purification system are effective qutrits. This suggests us an strategy to prepare highly dissonant states of two qubits: the preparation of pure entangled two-qutrit state, where the first qutrit should correspond to a two-qubit system in a Hilbert space of dimension three.

\subsection{Cavity assisted generation}

\begin{figure} 
\includegraphics[width=88mm]{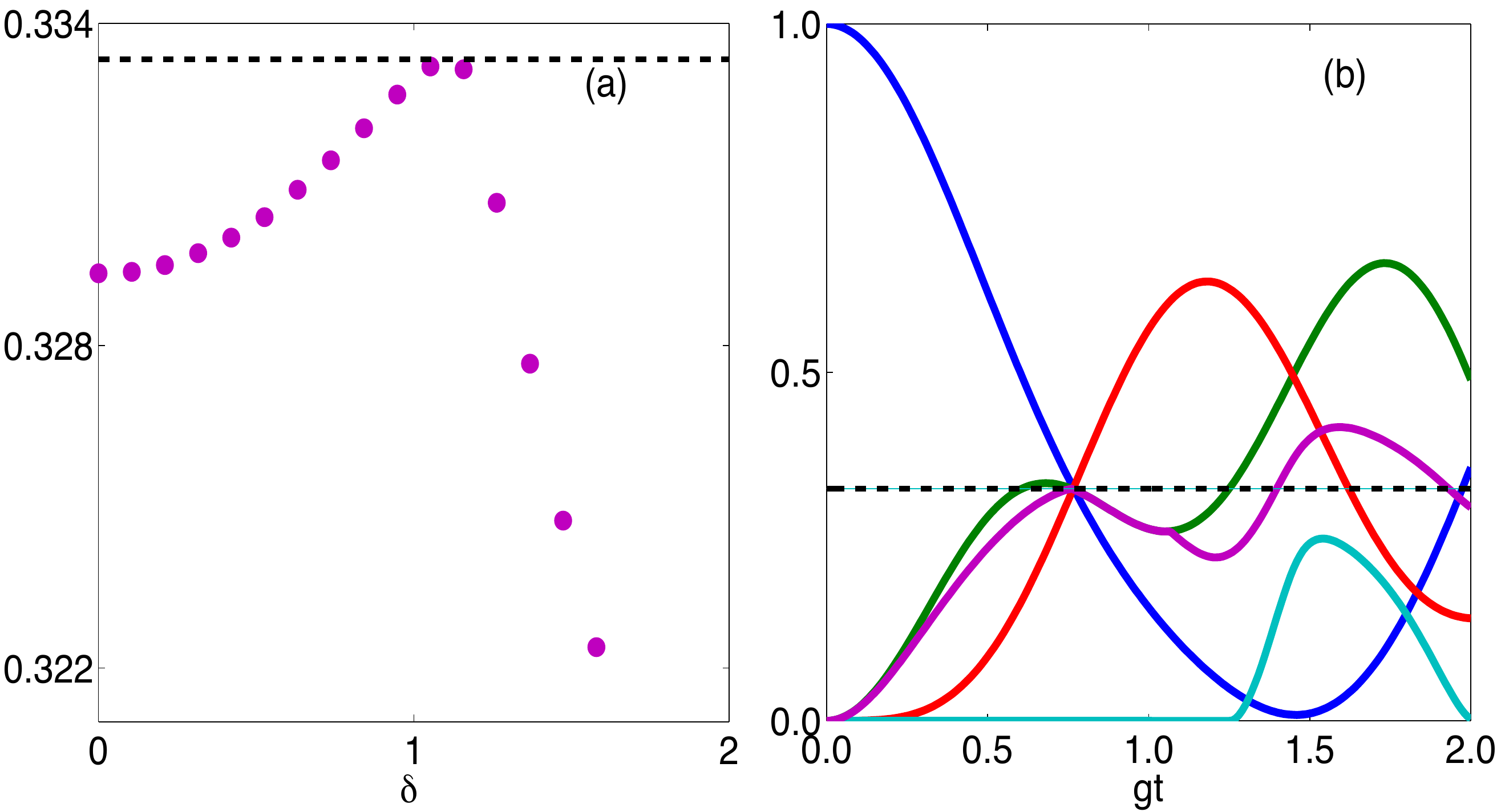} 
\caption{(a) Maximum quantum discord $Q$ (purple dots) as a function of the detuning $\delta.$ Black dashed line shows the maximum posible discord $Q_{max}=1/3.$ (b) Evolution of populations $|a_1 (t)|^2$ (blue line), $|a_2 (t)|^2$ (red line), $|a_3 (t)|^2$ (green line), quantum discord $Q(t)$ (purple line) and concurrence $C(t)$ (cian line) for $n=0$ for detuning $\delta=1.07g$. }
\label{fig2} 
\end{figure}

A physical model where we can apply this strategy is the one consisting in two two-level atoms interacting resonantly with a single mode of the electromagnetic field in the Tavis-Cummings model~\cite{tavis}. The Hamiltonian for such system is given by 
\begin{equation}
H=\hbar g \left ( a^\dagger J e^{i\delta t} + a J^\dagger e^{-i\delta t}\right) \label{ham1}
\end{equation}
where $J= \sigma_1 +\sigma_2$, with $\sigma_j = |g\rangle_j \langle e|$ and $\delta$ is the frequency difference between the field mode and the atomic transition frequency. If we assume both atoms initially in the excited state, that is $|ee\rangle $, and the field in a Fock state of $n$ excitations $|n\rangle$, the evolution leads to the state:
\begin{equation}
|\Psi(t)\rangle = a_1 (t) |ee\rangle |n\rangle+a_2 (t) |+\rangle |n+1\rangle+a_3 (t) |gg\rangle |n+2\rangle \label{psi1}
\end{equation}
with $|+\rangle = (|eg\rangle+|ge\rangle)/\sqrt{2}$. This state corresponds to an entangled state of qutrits:  the electromagnetic field mode lives within a three-dimensional Hilbert space spanned by the states $\{|n\rangle,|n+1\rangle,|n+2\rangle\}$. On the other hand, the atomic populations oscillates also within a thee-dimensional Hilbert space $\{|ee\rangle,|+\rangle,|gg\rangle\}$. In consequence, the dynamics provided by Hamiltonian Eq. (\ref{ham1}) results in an entangled state of two qutrits and then a highly dissonant state of two qubits may be prepared from this state. To see this, consider only the atomic part of the system by tracing out the bosonic mode: $\rho_{\rm at} = {\rm Tr}_{\rm field} (|\Psi(t)\rangle \langle\Psi(t) |)$. From Eq.~(\ref{psi1}), we have that
\begin{equation}
\rho_{\rm at} = |a_1 (t)|^2 |ee\rangle\langle ee |+|a_2(t)|^2 |+\rangle\langle + |+|a_3 (t)|^2 |gg\rangle\langle gg |\label{dicke1}
\end{equation}
This state has the same structure of state  in Eq.~(\ref{maximal}). In Fig.~\ref{fig2}(a) we show the maximal discord we found as a function of the detuning $\delta$ for the case $n=0$. We observe that when $\delta \sim 1.07 g$ the discord reaches approximately its maximum value of 1/3. In Fig.~\ref{fig2}(b) we show evolution of populations $|a_1 (t)|^2$, $|a_2 (t)|^2$ and $|a_3 (t)|^2$ for $n=0$ and $\delta=1.07g$. We observe that when $gt \sim 0.75$ all amplitudes in~(\ref{dicke1}) are approximately equals: $|a_1 (t)|^2  \approx |a_2 (t)|^2  \approx |a_3 (t)|^2 \approx 1/3$ and in consequence quantum discord reaches a maximum value (also $\approx$ 1/3). 

It is not difficult to see that when no detuning is considered, the quantum discord still reaches a high value  $(\sim 0.3329)$ in the absence of entanglement. However, when $\delta \neq 0 $, as the situation shown, the amount of discord increase until its maximum value in the absence  entanglement. Later entanglement  appears suddenly leading to higher values for discord.

\subsection{Off resonant collective atomic interaction assisted generation}

Another scenario where dissonant states can be  prepared, is when we consider the Tavis-Cummings model with two sets of atoms: a first set of two atoms coupled to the bosonic mode with strength coupling constant $g_1$  and a second set of $N$ atoms coupled to the same field mode with coupling constant $g_2$. All the $N+2$ atoms are coupled far from resonance to the cavity mode. In this case, the field mode is only virtually populated and the dynamics can be described by the effective Hamiltonian $H = H_0 + H_1 $~\cite{lopez}, where
\begin{eqnarray}
H_0 &=&  \hbar \lambda_1 J_1^\dagger J_1 + \hbar \lambda_2 J_2^\dagger J_2 \\
H_1 &=& \hbar\Omega_{12} \left (J_{1}^\dagger J_{2}e^{-i \delta t}+J_{2}^\dagger J_{1}e^{i \delta t} \right) \label{Heff}
\end{eqnarray}
with $\delta= \Delta_2-\Delta_{1}$, $J_{1}= \sigma_1 +\sigma_2$, $J_{2}= \sum_{j=3}^{N+2} \sigma_j $, and
\begin{eqnarray*}
 \\
\Omega_{12} &=& \frac{g_1 g_{2}}{2}\left(\frac{1}{\Delta_1}+\frac{1}{\Delta_{2}}\right) 
\end{eqnarray*}
where $\Delta_j$ are  the detuning between the $j$-th set of atoms and the field mode, respectively.

\begin{figure}[t]
\includegraphics[width=50mm]{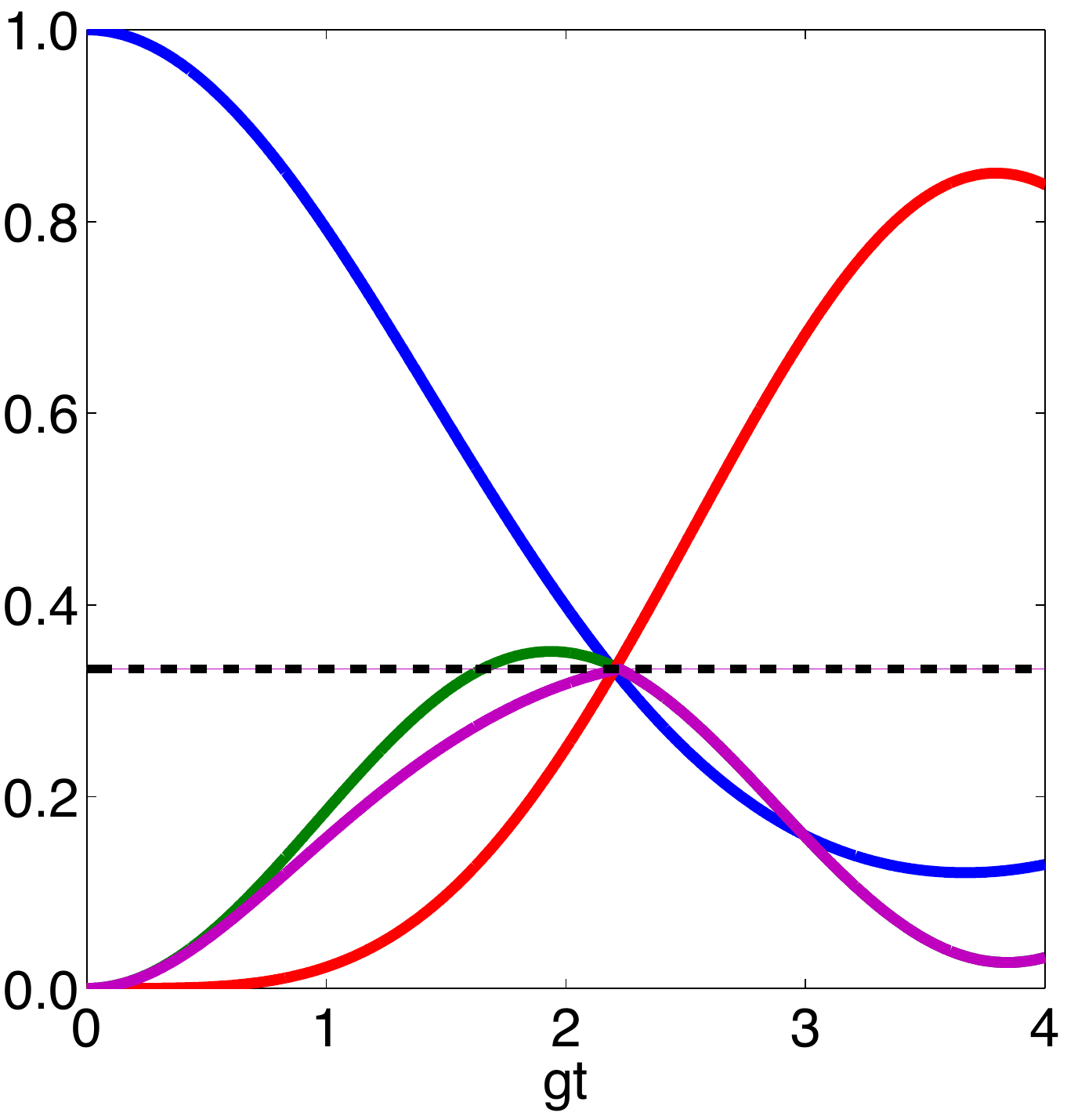} 
\caption{(a) Evolution of populations $|b_1 (t)|^2$ (blue line), $|b_2 (t)|^2$ (green line), $|b_3 (t)|^2$ (red line) and quantum discord $Q(t)$ (purple line) for $\Delta_1 = 11.2 g $, $\Delta_2 =10g$ ($\delta=-1.2g$) and $N=13$. Black dashed line shows the maximum posible discord without entanglement $Q_{max}=1/3.$}
\label{fig4ab}  
\end{figure}

If the atomic subsystems are initially in the symmetric subspace, the effective Hamiltonian (\ref{Heff}) will generate entangled states between these symmetric states only.  The first subsystem of two atoms behaves as an effective qutrit $\{|ee\rangle,|+\rangle,|gg\rangle\}$. The second set of $N$ atoms will be  described by the symmetric states with $k$ excitations $|D_k\rangle$.   We will have a two  qutrit only  if the number  of excitations in the system is two. This is true if we consider the initial state:
\begin{equation}
|\Psi(0)\rangle = |ee\rangle | D_0^N \rangle
\end{equation}
In such case, the dynamics is restricted to the subspace $\{|ee, D_0^N \rangle,|+, D_1^N \rangle,|gg, D_2^N \rangle\}$, where both atomic subsystems behaves as qutrits. The hamiltonian in this case can be rewritten as 
\begin{eqnarray}
H &=& \hbar \Omega_{12} (\sqrt{2N} |ee, D_0^N \rangle \langle +, D_1^N| e^{i\delta_a t}\notag \\ 
&& +\sqrt{4(N-1)} |+, D_1^N \rangle \langle gg, D_2^N| e^{i\delta_b t})+ {\it h.c.} 
\label{hdicke}
\end{eqnarray}
where,
\begin{eqnarray*}
\delta_a &=&  -\delta - \lambda_1- \lambda_2 (N-1)\\
\delta_b &=&  -\delta + \lambda_1- \lambda_2 (N-3)\\
\end{eqnarray*}
One of the advantages of considering different couplings and detunings is that we can set into resonance determined transitions~\cite{lopez,lopez1,lopez2}. For example, in this case we look for different values for $\Delta_1$, $\Delta_2$  and $N$ such that maximum discord could be reached.  In Fig. \ref{fig4ab}, we show the evolution of the populations $|b_j (t)|^2$ for $\Delta_1 = 11.2 g $, $\Delta_2 =10g$ and $N=13$. In the figure it can be observed that the quantum discord reaches an approximately maximum value ($\sim 0.333$) at $gt \sim 2.22$.

\section{Maximally correlated qutrits with no entanglement}

\begin{figure}[t]
\includegraphics[width=50mm]{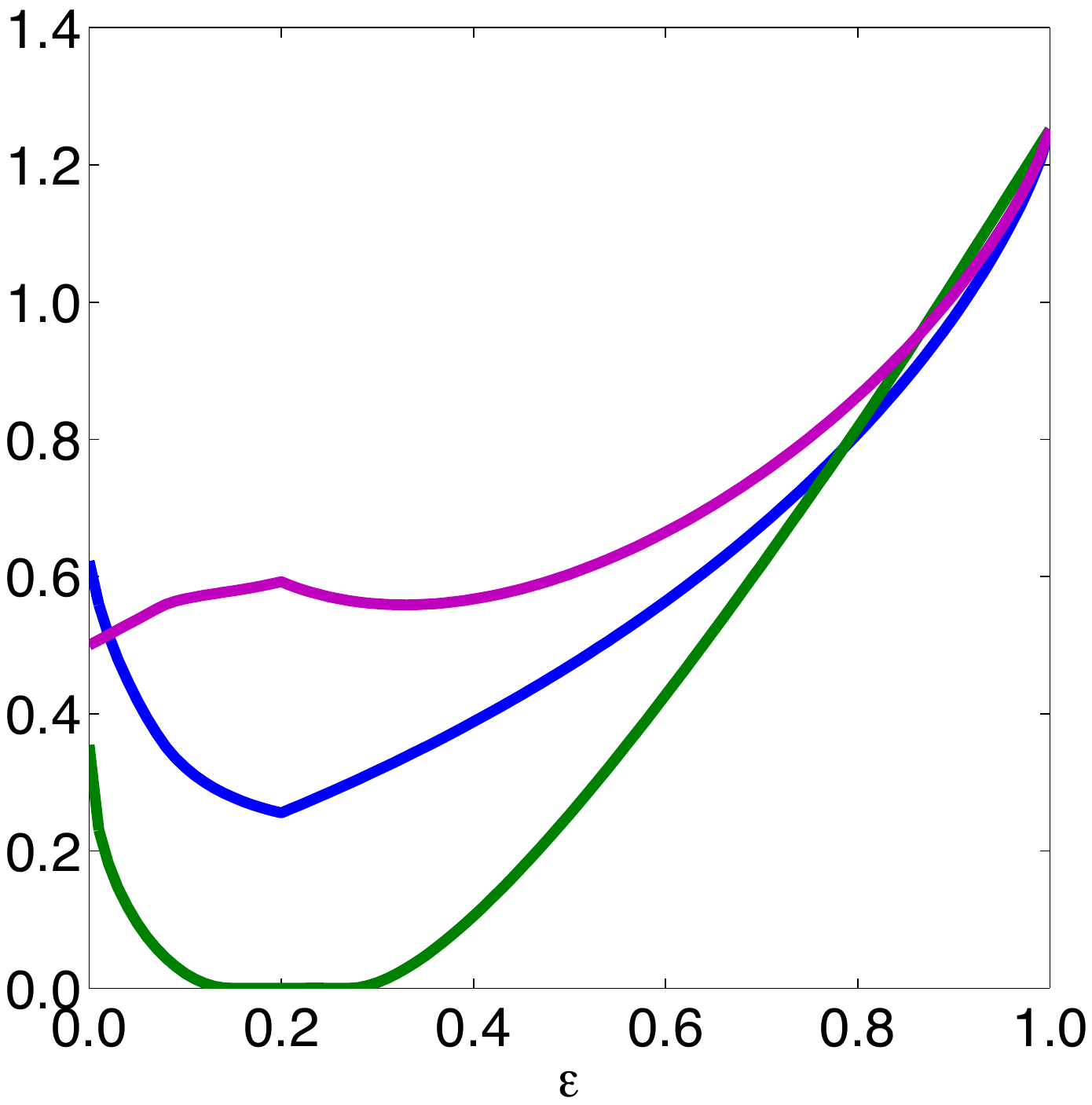} 
\caption{Evolution of classical correlations ${\cal{C}}(t)$ (blue line), quantum discord $Q(t)$ (purple line) and entanglement of formation (green line) as a function of the parameter $\epsilon$ for the state (\ref{rhoep}).}
\label{fig5}  
\end{figure}

Following the discussion, we can now study if the existence of these class of quantum states with maximal quantum correlation without entanglement are an exclusive property of bipartite states of qubits. To answer this question  Let us consider the simple generalization of the state (\ref{maximal}) for two pair of atoms in the symmetric space
\begin{eqnarray}
\rho=\sum_{k=0}^{4} d_k|D_k\rangle \langle D{_k}|
\label{rhoep}
\end{eqnarray}
where $|D_{k}\rangle$ are the symmetric Dicke state of four atoms with $k$ excitations. This state can be viewed as a correalted state of two qutrits $\{|ee\rangle,|+\rangle,|gg\rangle\}$.  In general this state will exhibit entanglement depending on the values of $d_k$, as happen for the state (\ref{rank3}). For example consider the simplest situation $d_2=\epsilon$, and  $d_{0}=d_{1}=d_3=d_4=1-\epsilon$. As is shown in Fig. \ref{fig5}, quantum correlations exhibit a maximum value for $\epsilon =1/5$ in the region where there is no entanglement. This state correspond to the generalization of state (\ref{maximal}) to the case of two qutrits, that is a flat  distribution in the symmetric subspace of two qutrits . 

As we are concern about the generation of correlated states in the absence of entanglement, let us consider the dynamics under Hamiltonian (\ref{ham1}), from which the state (\ref{rhoep}) can be generated from an initial condition of the form:
\begin{equation}
|\psi_0 \rangle = |D_4 \rangle |0\rangle \equiv |eeee\rangle |n\rangle
\end{equation}
that is,  all atoms in the excited state and the quantum field in the vacuum state.  The evolution will let us with

\begin{eqnarray} 
|\Psi(t)\rangle &=& a_1 (t) |D_4\rangle |0\rangle+a_2 (t) |D_3\rangle |1\rangle+ a_3 (t) |D_2\rangle |2\rangle \notag \\
&&+a_4 (t) |D_1\rangle |3\rangle+a_5 (t) |D_0\rangle |4\rangle. \label{psi4}
\end{eqnarray}

\begin{figure}[t]
\includegraphics[width=50mm]{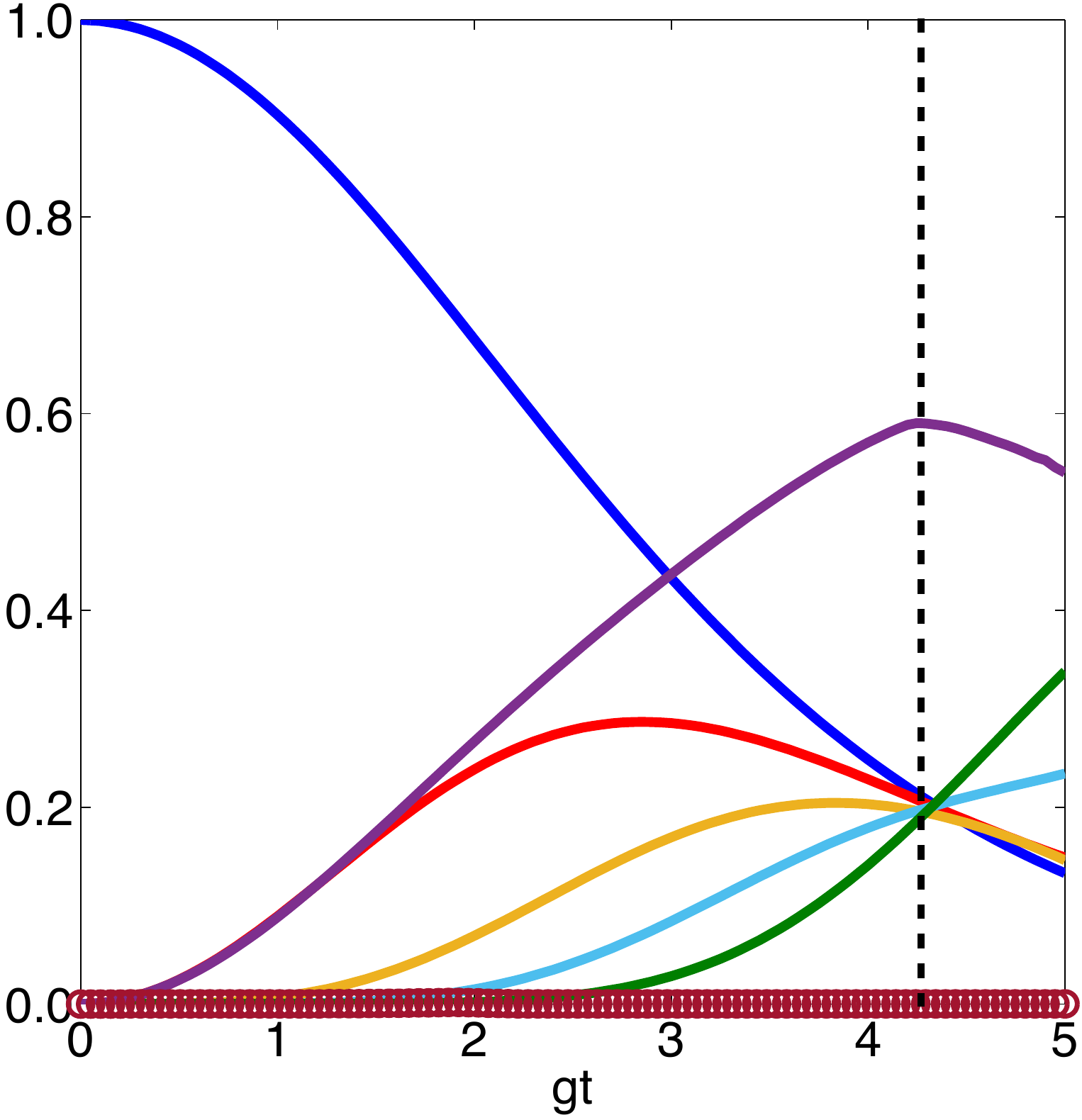} 	
\caption{Evolution of populations $|a_1 (t)|^2$ (blue line), $|a_2 (t)|^2$ (red line), $|a_3 (t)|^2$ (yellow line), $|a_4 (t)|^2$ (cian line) and $|a_5 (t)|^2$ (green line), quantum discord $Q(t)$ (purple line) and entanglement of formation (brown circles) for $\delta = 1.08 g$. Black dashed line shows the time when discord reaches its maximum value.}
\label{fig6}  
\end{figure}

Tracing out the bosonic mode, Eq.~(\ref{psi4}) will take the same form that Eq.~(\ref{rhoep}). As in the previous cases, the maximum quantum correlations in the absence of entanglement occurs when all probabilities are approximately equals, that is, $|a_j(t)|^2\approx 1/5$ . In Fig.~\ref{fig6} the evolution of the probabilities $|a_j (t)|^2$ is shown for $\delta = 1.08 g$. We see in this Fig. that quantum discord reaches a maximum value while the entanglement of formation is found to be zero in this time period. 

To calculate entanglement of formation and quantum discord for the bipartite states of qutrits showed in Fig. \ref{fig5} and Fig. \ref{fig6}, we have used the Simulated Annealing Algorithm (SAA) \cite{SAA}. The entanglement is carried out by searching for all pure state decomposition of the density matrix that minimize $\sum_{i}q_{i}E(|\psi_{i}\rangle)$, where $E(|\psi_{i}\rangle)$ is the von Neumann entropy.  Such decompositions are found by considering a purification of the density matrix $\sum_{i}\sqrt{q_{i}} |\psi_k\rangle |e_k\rangle$ and searching for the unitary transformation ($\mathbb{I} \otimes U_E)$ that is acting on the purification space only. Fo the calculation of quantum discord  the optimization is carried out with respect to all possible measurement on subsystem $B$, which requires to cover all possible projections $\Pi_l=\mathbb{I}\otimes V_B| l \rangle \langle l| V_B^{\dagger}$ where $l=0,1,2$ and $V_B$ is a unitary $3 \times 3$ matrix. There is an specific unitary $V_B$ that optimize the conditional entropy, and then quantum discord.The sampling the space of unitary transformations is carried out by changing the annealing parameter as $C = 10^{-9}e^{-2k}$, where $k = [1,K]$, and $K$ is the number of annealing processes. For every figure in this work we used $K = 10$, with $5 \times 10^4$ iterations for each $k$. The dimension of the purification space $M$ was 10.

In summary, we have investigated the generation of maximally dissonant bipartite states. These states has the interesting property that they hold the maximum possible quantum correlations without showing entanglement. To generate such states, we have proposed a theoretical method consisting in the generation of a maximally entangled state of two qutrits. One of the qutrits corresponds to the effective representation of a two-qubit system and the second qutrit corresponds to an ancillary system. This ancillary system could correspond to an effective description of a reservoir, such is the case of a our first example where a pair of impurity-bound electrons interact with a bath of conduction band electrons in a semiconductor. In the second example we use a bosonic mode as the ancillary system in the Tavis-Cummings model. In the third case we consider a bath of atoms as ancillary systems. In all cases we have addressed we find that maximally correlated states arise in the absence of entanglement. We finally extended the study searching for maximally correlated states without entanglement in higher dimensions. These finding shows that the existence of maximally correlated states in the absence of entaglement  is not an exclusive property of qubits.
\begin{acknowledgments}
Authors acknowledge financial support from CONICYT, DICYT 041631LC, Fondecyt 1161018, Fondecyt 1140194 and Financiamiento Basal FB 0807 para Centros Cient\'ificos y Tecnol\'ogicos de Excelencia.
\end{acknowledgments}


\begin{thebibliography}{99}
\bibitem{horo1} R. Horodecki, P. Horodecki, M. Horodecki, and K. Horodecki, Rev. Mod. Phys. \textbf{81}, 865 (2009).
\bibitem{Geza} O. G\"uhne and G. T\'oth, Phys. Rep. \textbf{474}, 1 (2009).
\bibitem{brasil1}S. P. Walborn, P. H. Souto Ribeiro, L. Davidovich, F. Mintert, and A. Buchleitner, Nature (London) \textbf{440}, 1022 (2006).
\bibitem{brasil2} M. P. Almeida, F. de Melo, M. Hor-Meyll, A. Salles, S. P. Walborn, P. H. Souto Ribeiro, and L. Davidovich, Science \textbf{316}, 579 (2007).
\bibitem{laurat}J. Laurat, K. S. Choi, H. Deng, C. W. Chou, and H. J. Kimble,
Phys. Rev. Lett. \textbf{99}, 180504 (2007).
\bibitem{review} K. Modi, A. Brodutch, H. Cable, T. Paterek, and V. Vedral,
Rev. Mod. Phys. \textbf{84}, 1655 (2012).
\bibitem{discord}H. Ollivier and W. H. Zurek, Phys. Rev. Lett. \textbf{88}, 017901 (2001); L. Henderson and V. Vedral, J. Phys. A: Math. Gen. \textbf{34}, 6899 (2001).
\bibitem{modi}K. Modi, T. Paterek, W. Son, V. Vedral, and M. Williamson,
Phys. Rev. Lett. \textbf{104}, 080501 (2010).
\bibitem{DQC1} E. Knill and R. Laflamme, Phys. Rev. Lett. \textbf{81}, 5672 (1998); A. Datta, A. Shaji, and C. M. Caves, Phys. Rev. Lett. \textbf{100}, 050502 (2008);
B. P. Lanyon, M. Barbieri, M. P. Almeida, and A. G. White, Phys. Rev. Lett. \textbf{101}, 200501 (2008).
\bibitem{lroa} L. Roa, J. C. Retamal, and M. Alid-Vaccarezza, Phys. Rev. Lett. \textbf{107}, 080401 (2011).
\bibitem{galve}F. Galve, G. L. Giorgi, and R. Zambrini, Phys. Rev. A \textbf{83}, 012102 (2011); F. Galve, G. L. Giorgi, and R. Zambrini, Phys. Rev. A \textbf{83}, 069905 (2011). 
\bibitem{ali}M. Ali, A. R. P. Rau, and G. Alber, Phys. Rev. A \textbf{81}, 042105 (2010); M. Ali, A. R. P. Rau, and G. Alber, Phys. Rev. A \textbf{82}, 069902 (2010).
\bibitem{sipe} K. S. Virk and J. E. Sipe, Phys Rev. B \textbf{72}, 155312  (2005). 
\bibitem{altintas1} F. Altintas, A. Kurt, and R. Eryigit, Phys. Lett. A \textbf{377}, 53 (2012); F. Altintas and R. Eryigit, Phys. Lett. A \textbf{376}, 1791 (2012).
\bibitem{tavis} M. Tavis and F. W. Cummings, Phys. Rev. \textbf{170}, 379
(1968).
\bibitem{lopez} C. E. L\'opez, F. Lastra, G. Romero, E. Solano, and J. C. Retamal, Phys. Rev. A \textbf{85}, 032319 (2012).
\bibitem{lopez1} C. E. L\'opez, J. C. Retamal, and E. Solano, Phys. Rev. A \textbf{76},
033413 (2007).
\bibitem{lopez2} L. Lamata, C. E. L\'opez, B. P. Lanyon, T. Bastin, J. C. Retamal, and E. Solano, Phys. Rev. A \textbf{87}, 032325 (2013).
\bibitem{SAA} S. Allende, D. Altbir, and J. C. Retamal, Phys. Rev. A \textbf{92}, 022348 (2015).
\end{thebibliography}
\end{document}